\documentclass[prl,twocolumn,nofootinbib,aps,preprintnumbers,amsmath,amssymb,superscriptaddress]{revtex4-2}
\usepackage{graphicx}
\usepackage{color}
\usepackage{bm}
\usepackage{amsmath}
\usepackage{xspace}
\usepackage[normalem]{ulem}
\usepackage{units}
\usepackage{dcolumn}
\usepackage{paralist}
\usepackage{natmove}
\usepackage{natbib}

\usepackage{hyperref}

\begin{document}

\title{First-Principles Prediction of Electronic Transport in Experimental Semiconductor Heterostructures via Physics-Based Machine Learning}
\author{Artem~K~Pimachev}
\affiliation{Ann and H.J. Smead Aerospace Engineering Sciences, University of Colorado Boulder, Boulder, Colorado 80303, USA}
\author{Sanghamitra~Neogi}
\email{sanghamitra.neogi@colorado.edu}
\affiliation{Ann and H.J. Smead Aerospace Engineering Sciences, University of Colorado Boulder, Boulder, Colorado 80303, USA}

\date{\today}

\begin{abstract}
First-principles techniques for electronic transport property prediction have seen rapid progress in recent years. However, it remains a challenge to model heterostructures incorporating variability due to fabrication processes. Machine-learning (ML)-based materials informatics approaches (MI) are increasingly used to accelerate design and discovery of new materials with targeted properties, and extend the applicability of first-principles techniques to larger systems. However, few studies exploited MI to learn electronic structure properties and use the knowledge to predict the respective transport coefficients. In this work, we propose an electronic-transport-informatics (ETI) framework that trains on {\em ab initio} models of small systems and predicts thermopower of silicon/germanium heterostructures beyond the length-scale accessible with first-principles techniques, matching measured data. We demonstrate application of MI to extract important physics that determines electronic transport in semiconductor heterostructures, breaking from combinatorial strategies pursued especially for thermoelectric materials. We anticipate that ETI would have broad applicability to diverse materials classes.
\end{abstract}

\maketitle

Semiconductor heterostructures have brought about tremendous changes in our everyday lives in the form of telecommunication systems utilizing double-heterostructure lasers, heterostructure light-emitting diodes, or high-electron-mobility transistors used in high-frequency devices, including satellite television systems~\cite{alferov2001nobel}. Silicon (Si)/germanium (Ge) heterostructures, in particular, have emerged as key materials in numerous electronic~\cite{thompson200490,meyerson1994high,nissim2012heterostructures,paul2004si}, optoelectronic~\cite{koester2006germanium,liu2010ge,tsaur1994heterojunction}, and thermoelectric devices~\cite{alam2013review,taniguchi2020high}, and promising host of spin qubits~\cite{shi2011tunable}. Recent developments of nanofabrication and characterization techniques achieved great control over the growth of Si/Ge heterostructures~\cite{euaruksakul2013heteroepitaxial,brehm2017site,lee2019interplay,chen2013role,david2018new}. Nevertheless, fabrication of heterostructures is strongly affected by strain relaxation in component layers~\cite{paul2004si}, and the resulting electronic properties show high variability due to inconsistent fabrication dependent structural parameters~\cite{samarelli2013thermoelectric,taniguchi2020high,koga2000experimental,taniguchi2020high}. A few theoretical studies discussed the effect of non-idealities on electronic properties of heterostructures~\cite{watling2011study,vargiamidis2019hierarchical}, however, these studies were parametric in nature. It is essential to acquire a comprehensive understanding of the complex relationship between growth dependent parameters and electronic properties, to attain targeted semiconductor heterostructure design with reliable electronic performance. {\em Ab initio} techniques enable prediction of materials properties with minimal experimental input, however, often come with large computational costs. In particular, the calculations of electronic transport coefficients (such as, thermopower or conductivity) require large number of individual energy calculations and computational costs can accrue quickly. It remains a challenge to model electronic transport coefficients of technologically relevant heterostructures incorporating full structural complexity, representing the vast fabrication dependent structural parameter space. 

Recent studies demonstrated the ability of data driven techniques to predict the results of new calculations at little additional computational cost, using previous {\em ab initio} model data as input~\cite{ramprasad2017machine,schleder2019dft}. The use of machine learning (ML) models showed remarkable successes in accelerating atomistic computations and extending applicability of first-principles techniques to predict properties of larger systems~\cite{snyder2012finding,behler2016perspective,xia2018quantum,gong2019predicting}. Machine-learning-based materials informatics (MI) are increasingly being used to accelerate design and discovery of new materials and structures~\cite{ward2016general,meredig2014combinatorial,jain2016computational}, facilitated by large amounts of data generated with high-throughput density functional theory (DFT) calculations~\cite{schleder2019dft} or available through databases~\cite{Jain2013,Draxl2019TheNL}. Most of these studies aimed at identifying structures that optimize the target property. ML models are being used to explore the relationship between structure and electronic transport property, especially in the context of thermoelectric materials. However, the focus remained on combinatorial approaches to identify compounds that optimize the relevant electronic transport coefficients, such as thermopower~\cite{iwasaki2019machine} or electronic power factor~\cite{hou2019machine}. Only recently, some attempts have been made to use MI to learn and predict atomic scale dynamics~\cite{xie2019graph}. Few studies exploited ML techniques to establish relationship between the electronic structure properties and the respective transport coefficients~\cite{li2020neural,lopez2014modeling}. A vast amount of information is generated during a single {\em ab initio} electronic structure property calculation. Therefore, there is great benefit to develop approaches that can harvest information from previous calculations to predict properties of new systems, a priori.

In this work, we propose a first-principles-based electronic-transport-informatics (ETI) framework that is trained on the electronic structure properties of small systems and predicts transport coefficients, namely the thermopowers of experimental semiconductor heterostructures. The framework is built on the hypothesis that functional relationships between local atomic configurations, $CN({\bf r})$, and their contributions to global energy states, $E$, are preserved when the configurations are part of a nanostructure with different composition and/or dimensions. The rationale for the hypothesis is rooted in the fundamental insight that material's physical properties, ranging from mechanical to electronic, are intimately tied to the underlying symmetry of the crystal structure~\cite{nye1985physical}. This conjecture allows one to probe the local configuration-energy state relationships, $f(CN({\bf r}),E)$, in few-atom fragment units with varied atomic environments, and harness the information to predict $f(CN({\bf r}),\hat{E})$'s that develop in larger nanostructures, hosting similar local environments, $CN({\bf r})$. We implement this hypothesis to extrapolate the insight acquired from small {\em ab-initio} models to predict local $f(CN({\bf r}),\hat{E})$'s in experimental semiconductor heterostructures. We estimate the global energy states of the heterostructure with known $CN({\bf r})$'s by assimilating the predicted $f(CN({\bf r}),\hat{E})$'s. The energy states are then used to predict Seebeck coefficients ($S$) or thermopowers, that are validated against numerical results obtained with first-principles methods (DFT), or experimental data. We anticipate that the hypothesis can be similarly applied to obtain other electronic transport coefficients. Our ETI framework thus establishes that MI can be exploited to address the gap between ideal {\em ab-initio} models and systems realized with nanofabrication techniques.

\section{Results and Discussion}

\begin{figure*}
\begin{center}
\includegraphics[width=1.0\linewidth]{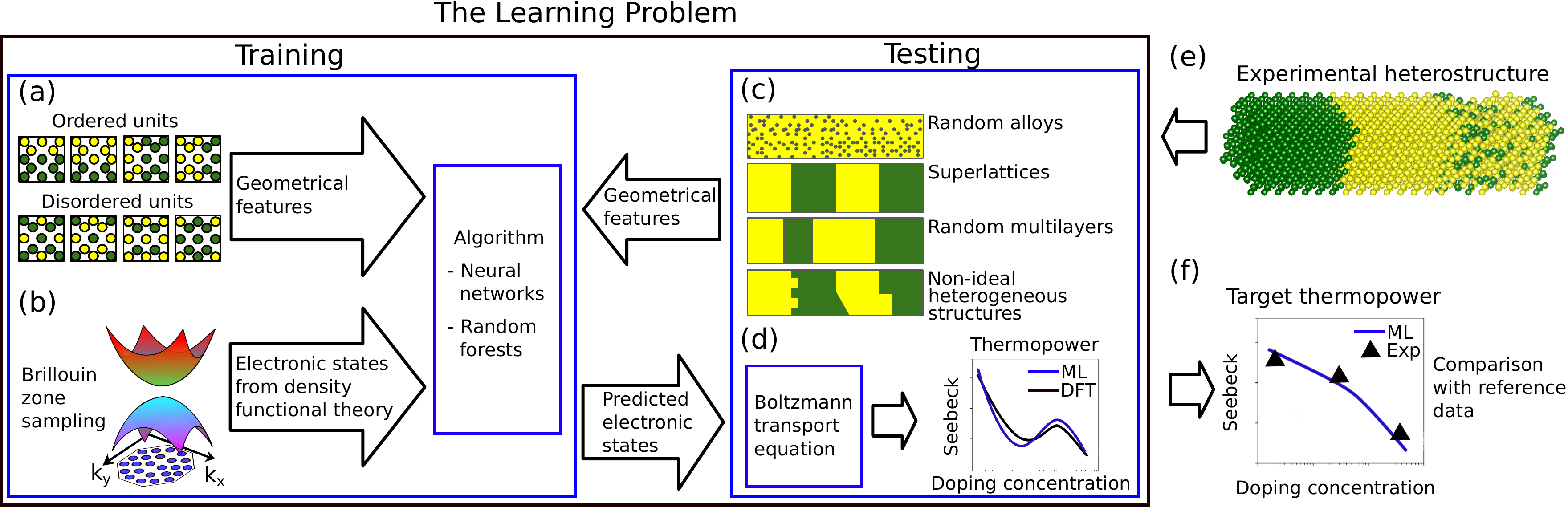}
\setlength{\belowcaptionskip}{-25pt} 
\caption{{\bf Workflow outline to predict electronic transport properties of experimental heterostructures using machine learning (ML) algorithms:}
(e) Representative configuration of an experimental heterostructure. (a) Ordered and disordered heterostructure fragment units of varied compositions. Several features describing local atomic environment, $CN({\bf r})$, of these units are used to train ML algorithms. (b) Property value corresponding to features: electronic structure properties, $E(k,b)$, computed with density functional theory (DFT), where $b$ is band index and $k$-points sample Brillouin zones of the units. Neural networks and random forests algorithms are used to learn structure-property relationship, $f(CN({\bf r}),E(k,b))$ of the units. (c) ML models are tasked to predict $\hat{E}(k,b)$'s of larger structures of varied compositions. (d) Cross-validation is carried out by comparing thermopower or Seebeck coefficients ($S$) calculated from $\hat{E}(k,b)$ and direct DFT results. $S$ is computed using semi-classical Boltzmann transport equation (BTE). (e) Finally, target $S$ of experimental heterostructures (e) is computed from $\hat{E}(k,b)$ using BTE, and compared with experimental data (f).}
\label{fig:ETI-model}
\end{center}
\end{figure*}

Figure~\ref{fig:ETI-model} shows the outline of the ETI framework that provides first-principles prediction of thermopowers of experimental Si/Ge heterostructures. The ML model learns the relationships between atomic configurations ($CN({\bf r})$) and electronic bands ($E$, panel (b)) present in 16-atom fragment training models (cartoons in panel (a)), and predicts environment-electronic state correlations, $f(CN({\bf r}),\hat{E})$, existing in larger heterostructures (panels (c, e)). The components of our ETI framework are: ({\bf \textit 1}) creation of a data resource to harvest $f(CN({\bf r}),E)$, ({\bf \textit 2}) formulation of a representation that can uniquely describe $CN({\bf r})$, and help characterize $f(CN({\bf r}),E)$, in a fragment unit or target heterostructure, ({\bf \textit 3}) choice of ML algorithms to discover correlations in training data, and, ({\bf \textit 4}) testing ML predictions for new structures against known data.

({\bf \textit 1}) {\em Creation of Data Resource:} We explored materials property databases, such as Material Project~\cite{Jain2013} and NOMAD Repository \& Archive~\cite{Draxl2019TheNL} to collect training data. However, the electronic structure property data of only limited number of Si/Ge structures are available. Additionally, the available data do not provide necessary sampling of the structure Brillouin zones (BZ) to converge electronic transport coefficients, requiring us to create our own data resource. In order to minimize data generation efforts, we limit the number of electronic property calculations, and instead, mine the large amount of information generated in individual calculations. In our recent publications, we presented extensive investigations of the electronic structure and transport properties of Si/Ge heterostructures~\cite{proshchenko2019heat, proshchenko2019optimization,settipalli2019theoretical,proshchenko2019modulation}. These past data and insights greatly facilitate the development of the ETI framework. Based on the acquired insights~\cite{proshchenko2019optimization,settipalli2019theoretical,proshchenko2019modulation}, we follow two strategies: ({\textit i}) careful choice of fragment training units, and (\textit{ii}) utilizing atomistic information generated from individual DFT calculations of the units as training data. 

The first strategy is implemented by choosing Si/Ge systems with varied strain environment as training fragments. The choice is guided by the fact that the bands of Si/Ge heterostructures are significantly affected by strain~\cite{proshchenko2019optimization,proshchenko2019modulation,settipalli2019theoretical,hinsche2012thermoelectric}. Application of strain led to more than an order of magnitude variation in electronic properties over the non-strained materials~\cite{peter2010fundamentals,ridley1999quantum,schaffler1997high}. In heterostructures, strain is generated due to various mechanisms including structural (lattice mismatch, presence of defects), thermal expansion or chemical (phase transition) changes. Naturally, the strain environment is variable, and the resulting electronic properties are unpredictable, making the problem ideal for the application of ML techniques. Panel (a) of Fig.~\ref{fig:ETI-model} shows cartoon representations of the two categories of the fragment training systems. The 16-atom models include ordered layered Si/Ge superlattices (SLs) and disordered Si-Ge ``alloys" (see Methods section for details). We acknowledge that the small size of the model units along with the imposed periodic boundary conditions do not reflect true randomized alloy configurations. Nevertheless, the models allow us to explore $f(CN({\bf r}),E)$ in these binary systems as a function of variable atomic environments. The remarkable successes shown by MI approaches using DFT generated data~\cite{ramprasad2017machine,schleder2019dft} inspired us to use DFT to generate training data. The DFT computed electronic structure properties and energy bands of the model units serve as training data and benchmark for cross-validation tests.

({\bf \textit 2}) {\em Formulation of Representation:} Identification of a minimal set of features is crucial to formulate relevant structure-property relationships~\cite{ramprasad2017machine,ward2017including,ghiringhelli2015big}. For our ETI framework, it is essential that the features describe sub-Angstrom-scale structural details because of the following reasons: ({\textit i}) electronic transport in a heterostructure is highly sensitive to local structural environment, and (\textit {ii}) success of ETI is based on the hypothesis that $f(CN({\bf r}),E)$ is preserved across structures hosting similar environment, and determines the electronic transport properties. Thus, it is essential that the correlations, $f(CN({\bf r}),E)$, are built upon fine details of $CN({\bf r})$, to ensure the transferability of the framework across structures. However, $f(CN({\bf r}),E)$ is expected to be multivariate and highly nonlinear. Hence, we are tasked to identify a feature subset that is strongly correlated with the electronic transport properties, from a large structural parameter space. A diverse set of elemental properties are used as features in MI~\cite{ward2017including}. Since our heterostructures are binary, the elemental-property-based features differ only slightly across the various configurations and are not expected to provide unique information to develop $f(CN({\bf r}),E)$. We consider only one elemental-property-based feature, computed from the electronegativity difference of the species (Si, Ge). Instead, we include multiple {\em global} and {\em local} structural features that are directly affected by strain. {\em Global} features include atomic composition of the systems (e.g., Ge concentration) and lattice constants ($a, \, b, \, c$).

\begin{figure}
\begin{center}
\includegraphics[width=\linewidth]{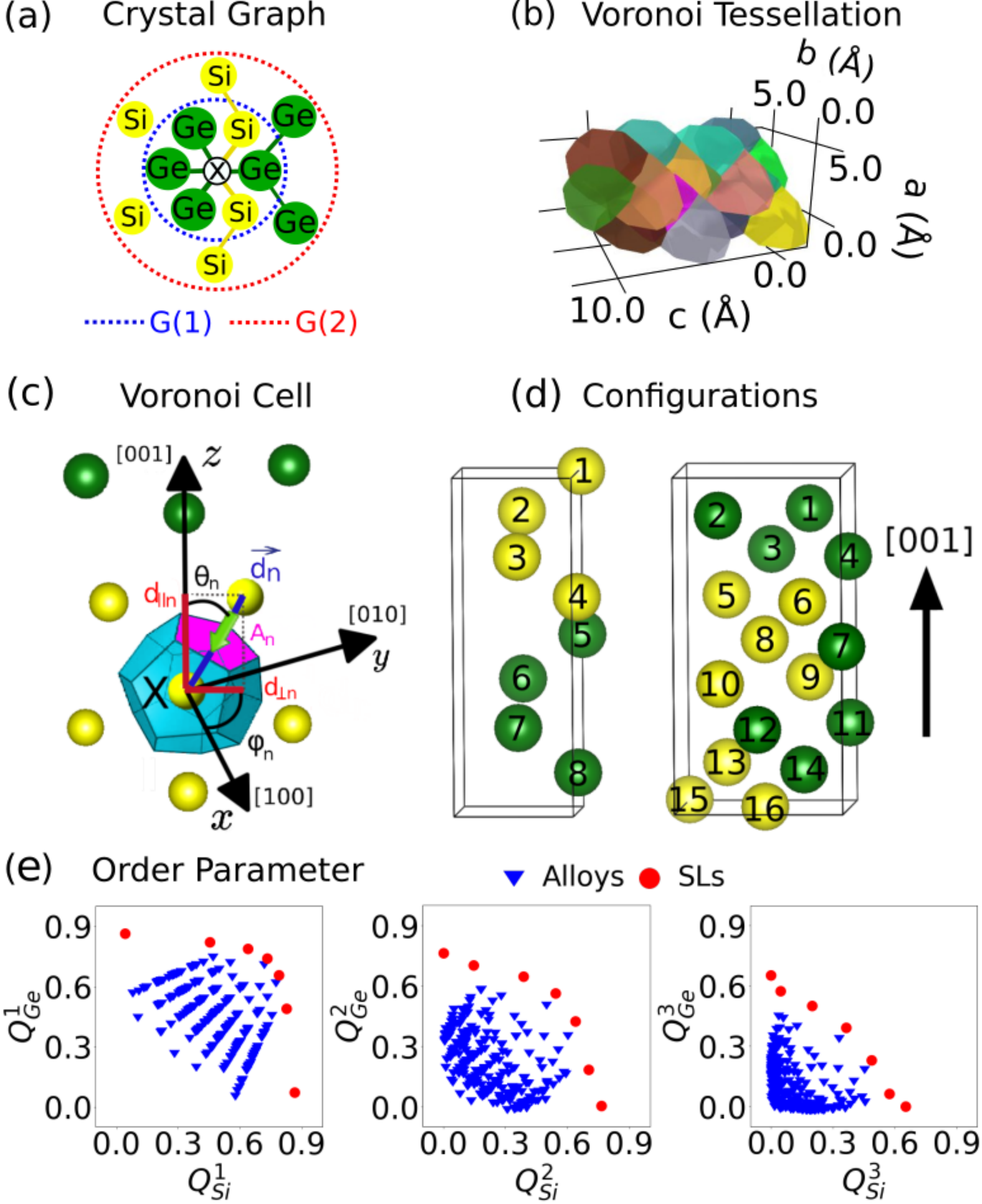}
\caption{{\bf Formulation of features using Voronoi tessellations:} (a) Representative neighbor crystal graphs connecting atom at node $X$ with $order =1$ ($G(1)$ (blue)) and $2$ ($G(2)$ (red)) neighbors. (b) Voronoi tessellations of a model Si$_4$Ge$_4$ SL. (c) Representative Voronoi cell in a SiGe model heterostructure. (d) Selected configurations to illustrate order parameter concept: (i) Si$_4$Ge$_4$ SL, (ii) Si$_8$Ge$_8$ disordered ``alloy". (e) Order parameters of ordered and disordered training units.}
\label{fig:crystal-graph}
\end{center}
\end{figure}

To determine the relevant {\em local} features describing $CN({\bf r})$, we express the structures as crystal graphs that encode both atomic information and bonding environments~\cite{xie2018crystal,xie2018hierarchical,gong2019predicting}. Crystal graph based ML models have shown great success in recent years for first-principles materials property prediction~\cite{ward2017including,xie2018crystal,xie2018hierarchical,gong2019predicting}. Figure~\ref{fig:crystal-graph}(a) shows a representative crystal graph $G$ of a typical SiGe configuration. The atom $X$ and the neighboring atoms form nodes, and the interatomic distances constitute the edges. We identify the neighboring atoms from Voronoi tessellations (VT) of the crystal structure. Figure~\ref{fig:crystal-graph}(b) shows VT of a model Si$_4$Ge$_4$ SL. The neighbors occupy adjoined cells and share faces in the tessellations. Thus, each face of a Voronoi cell marks a specific nearest neighbor of the selected atom. Figure~\ref{fig:crystal-graph}(c) shows a representative Voronoi cell in a representative Si/Ge configuration. The VT approach is particularly beneficial for our study since the tessellations are uniquely defined for a given local environment, and are insensitive to global dimensions of the structure. Therefore, VT-derived features help transfer $f(CN({\bf r}),E)$'s across structures of varying dimensions. VT-derived features has facilitated successful MI prediction of formation enthalpies~\cite{ward2017including}.
In total, we describe $CN({\bf r})$ of each configuration with 100 features. We have included extensive discussions of all the features in SI. The relative feature importance data shown in Fig.~S7 reflects the strong influence of VT-derived features on the performance of the ETI framework.

In our feature set, the local spatial ordering of atoms is represented by order parameters, $Q^{order}_{X}$~\cite{cowley1950approximate,ward2017including}, where 
\begin{equation}
  Q^{order}_{X} = \sum_{paths}\prod_{\substack{steps \ in \\ G(order)}}\underbrace{\frac{A_n\delta_{nX}}{\sum_{a}A_a-\sum_{b} A_b}}_{\text{fractional weight}}.  
\label{eq:QX}
\end{equation}
The paths connect all neighboring atoms up to a specified order in the crystal graph, e.g., $G(1)$ (blue), $G(2)$ (red) in Fig.~\ref{fig:crystal-graph}(a). We limit to order $ = 3$ ($G(3)$), because higher order graphs do not significantly impact the predictions, however, the cost to construct them increases proportionally with the volume containing the neighbors, $\sim $order$^3$. The Kronecker delta function in the numerator further restricts consideration of paths connecting only atoms of same type as $X$, yielding species aware crystal graphs. For example, the paths connecting the Si (yellow) or the Ge (green) circles in Fig.~\ref{fig:crystal-graph}(a) assume atom $X$ to be of type Si (yellow) or Ge (green), respectively. 
A typical step along a path connecting neighbors to $X$ is shown by the arrow (green) in Fig.~\ref{fig:crystal-graph}(c). The step crosses the face of index $n$ and area $A_n$ (magenta), normal to its direction. The ratio between area, $A_n$, and the sum over all areas the step could possibly cross, $A_a$, that are part of other non-backtracking paths, determines the fractional weight (Eq.~\ref{eq:QX}). Thus, the fractional weight of each step can be understood as the probability of taking the step. The product of fractional weights of all steps in a path determines the effective weight, the probability of choosing the path. The sum of the effective weights of all possible non-backtracking paths in $G(1), G(2) \ \text{and} \ G(3)$ results in $Q_X^{order}$. Figure~\ref{fig:crystal-graph}(e) shows the variations of $Q^{order=1,2,3}_{Si}$ and $Q^{order=1,2,3}_{Ge}$, for the 357 ordered (7) and disordered (350) fragment units. The scatter plot illustrates that we can distinguish each SiGe configuration by $Q^{order}_{Si}$ and $Q^{order}_{Ge}$. For example, the distinct clusters of data points representing the layered SL and the ``alloy" configurations can be noted. Thus, the order parameters are highly effective in distinguishing configurations with different degrees of structural ordering. From left to right in Fig.~\ref{fig:crystal-graph}(e), higher order parameter values decrease at a fast rate for disordered ``alloy" configurations, but only slightly for SLs. 

To further distinguish the anisotropic bonding environment of a SL compared to a disordered unit, we define directionally-biased order parameters, $Q^{\Omega=(x,y,z),order}_{X}$. The bias is implemented by calculating the fractional weights using projections of $A_n$ along a chosen direction only (see SI). In Table~\ref{table:TableQ1}, we show $Q^{\Omega,order}_{X}$'s of the atoms of the Si$_4$Ge$_4$ SL configuration shown in Fig.~\ref{fig:crystal-graph}(d). As a reference, these order parameters are all equal to $1$ for bulk systems. The in-plane order parameters, $Q^{x,order}_{X}$, $Q^{y,order}_{X}$, are equal, due to the rotational symmetry of the configuration around $z$-axis, aligned along [001]. In comparison, $Q^{z,order}_{X}$ is lower and decreases faster with the order number, reflecting heterogeneous stacking along $z$ direction. $Q^{z,order}_{X}$ can be used to identify the different atomic environments along $z$ direction, e.g., $Q^{z,1}\sim 0.5 - 0.6$ represents interface atoms and $Q^{z,1}\sim 0.9 - 1.0$ indicates inner atoms. The order parameter values are higher for inner atoms and lower for interface atoms. This is due to the presence of greater number of same species neighbors resulting in more paths contributing to order parameters of inner atoms. The order parameters also highlight the reflection symmetry with respect to the $x-y$ plane, yielding identical values for atom pairs such as $(1,2)\equiv(4,3)$ and $(5,6)\equiv(8,7)$. In comparison, the order parameters of the Si$_8$Ge$_8$ random ``alloy" configuration shown in Fig.~\ref{fig:crystal-graph}(d) do not show any specific pattern and decrease fast with the order number reflecting the disordered atomic arrangement (See SI Table III). In Fig.~S6, we show the species-aware, directionally-biased order parameters of all SL and ``alloy" training units. 
\begin{table}[ht]
\caption{Si$_4$Ge$_4$ SL order parameters}
\centering 
\begin{tabular}{c c c c c c c c c c c}
\hline\hline                  
Atom & \# & $Q^{x,1}$ & $Q^{y,1}$ & $Q^{z,1}$ & $Q^{x,2}$ & $Q^{y,2}$ & $Q^{z,2}$ & $Q^{x,3}$ & $Q^{y,3}$ & $Q^{z,3}$\\ [0.5ex]
\hline                  
Si&	1&  0.55&	0.55&	0.51&	0.49&	0.49&	0.45&	0.36&	0.36&	0.29\\
Si&	2&  0.95&	0.95&	0.89&	0.64&	0.64&	0.53&	0.43&	0.43&	0.34\\
Si&	3&  0.95&	0.95&	0.89&	0.64&	0.64&	0.53&	0.43&	0.43&	0.34\\
Si&	4&  0.55&	0.55&	0.51&	0.49&	0.49&	0.45&	0.36&	0.36&	0.29\\
Ge&	5&  0.53&	0.53&	0.49&	0.47&	0.47&	0.46&	0.37&	0.37&	0.30\\
Ge&	6&  0.97&	0.97&	0.94&	0.70&	0.70&	0.58&	0.47&	0.47&	0.36\\
Ge&	7&  0.97&	0.97&	0.94&	0.70&	0.70&	0.58&	0.47&	0.47&	0.36\\
Ge&	8&  0.53&	0.53&	0.49&	0.47&	0.47&	0.46&	0.37&	0.37&	0.30\\ [1ex]      
\hline
\end{tabular}
\label{table:TableQ1}
\end{table}
The order parameter feature $Q^{\Omega,order}_{X}$ is particularly important since directional ordering controls the atomic orbital contributions to energy bands in Si/Ge heterostructures~\cite{proshchenko2019modulation}. We have included further illustrations of the order parameter concept in SI.

\begin{figure*}
\begin{center}
\includegraphics[width=1.0\linewidth]{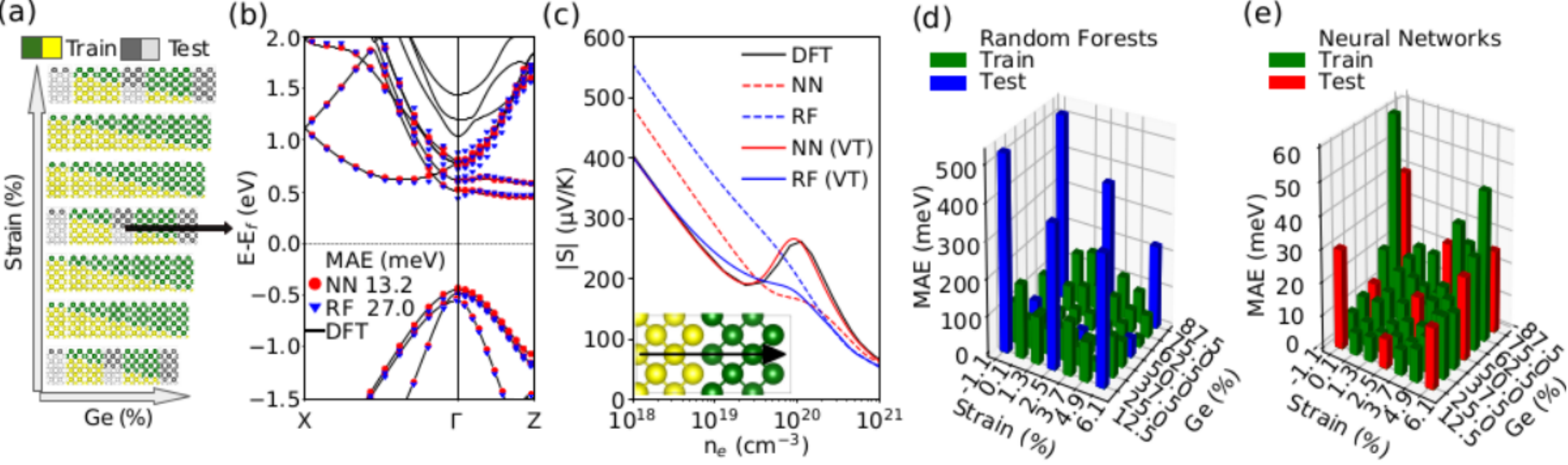}
\caption{{\bf Training and testing ML algorithms on ideal substrate strained layered structures:} (a) Representative configurations of 16-atom train and test units: ideal superlattices (SL) with varied compositions and external strain. Colored and gray cartoons represent train and test units, respectively. Each configuration is subjected to global in-plane substrate strains of varied magnitude. The unit structure in the middle correspond to a strain-symmetrized Si$_4$Ge$_4$ SL (see arrow). (b) Energy bands of strain-symmetrized Si$_4$Ge$_4$ SL calculated with DFT (black solid lines) and predicted by neural network (NN) (red circle) and random forests (RF) (blue inverted triangle) algorithms. (c) Thermopower calculated from bands obtained with DFT (black solid line) and predicted by NN (red) and RF (blue) algorithms. Solid and dashed lines represent predictions from ML models trained with and without inclusion of features from Voronoi tessellations (VT), respectively. (d,e) Mean absolute errors (MAE) of predicted energies of train and test structures using ML algorithms.}
\label{fig:strained_ideal_SL}
\end{center}
\end{figure*}

({\bf \textit 3}) {\em Choice of ML algorithm:}
We use supervised neural network (NN) and random forests (RF) algorithms to learn $f(CN({\bf r}),E)$, and compare performance of the two algorithms in predicting $\hat{E}$ for given $CN({\bf r})$ . 

\noindent \uline{NN Model:} Our model consists of an input layer consisting of 128 nodes, three fully connected hidden layers each with 256 nodes, and an output layer with nodes equal to the number of energy values, $\hat{E}(k,b)$: $k_x\times k_y \times k_z \times b$. We consider six valence and six conduction bands $(b)$, and a $21\times21\times21$ $k$-point mesh to sample the respective BZ, resulting in $21\times21\times21\times12$ $E$ values for each training configuration. We tested that such sampling of $E$-values yields necessary convergence of Seebeck coefficients~\cite{proshchenko2019optimization,proshchenko2019modulation} (Fig.~S9). The size of the model input is equal to the number of features considered. We use 100 features to describe each training unit and a batch size of 32 structures to train the model. Therefore, the NN algorithm is supplied with $32\times100$ input values at each iteration during training. The NN model is tasked to formulate $f(CN({\bf r}),E)$, relating features of $CN({\bf r})$ and the target electronic states $\hat{E}$, parametrized by weights $W$. The training is performed by iteratively updating the weights to minimize the mean absolute error (MAE) between the calculated and the predicted energies, 
\begin{equation}
MAE(E,\hat{E})=\frac{\sum_{k,b}|E(k,b)-\hat{E}(k,b)|}{(\text{number of $k$-points})(\text{number of bands})}.
\label{eq:MAE}
\end{equation}
We employ the ADAM stochastic optimization method for gradient descent to minimize the loss function (MAE).
The high-level NNs are implemented using the Keras library~\cite{chollet2015keras} written in Python. In all NN models, the Rectified Linear Unit activation functions are utilized. Five-fold cross-validation tests are performed to avoid overfitting. The optimized weights,
\begin{equation}
\underset{W}{min}~MAE(E,f(CN({\bf r});W)),
\end{equation}
are then used to predict $21\times21\times21\times12$ $\hat{E}$ values for unknown systems.

\noindent \uline{RF model:} RF models~\cite{breiman2014random} are shown to be computationally inexpensive and robust to overfitting of data~\cite{ward2017including}. The algorithm assembles results of several decision trees, each built from random selection of training data that include both features and example training energy values. The selected data is further partitioned into subsets based of decision rules that constitute branches of the tree. For example, the subsets can be formed based on order parameter values, e.g., $Q^{z,1}\sim 0.5 - 0.6$, representing different atomic environments (see Table~\ref{table:TableQ1}). The decision rules identify features that minimize the intrasubset variation of electronic energies. The leaves of the tree are then assigned to an energy value that maximizes fitting over the subset data. Such tree generation process is then repeated for other random subsets of training data. The final predictions are obtained by averaging the predicted energies over all trees. We implement the RF module available in the scikit-learn Python package~\cite{pedregosa2011scikit}. 
The input and output are identical to the ones used for the NN algorithm. We use 100 regression trees per ensemble and set all other parameters to default values recommended for the package. We did not observe any notable change in the predicted energy values by increasing the number of trees to 200 and 300.

In the following, we discuss the performance of the two ML algorithms in predicting electronic transport coefficients of three classes of SiGe heterostructures: (1) ideal strained superlattices, (2) non-ideal heterostructures with irregular layer thicknesses and imperfect layers, and (3) experimental heterostructures. 

\begin{figure}
\begin{center}
\includegraphics[width=0.9\linewidth]{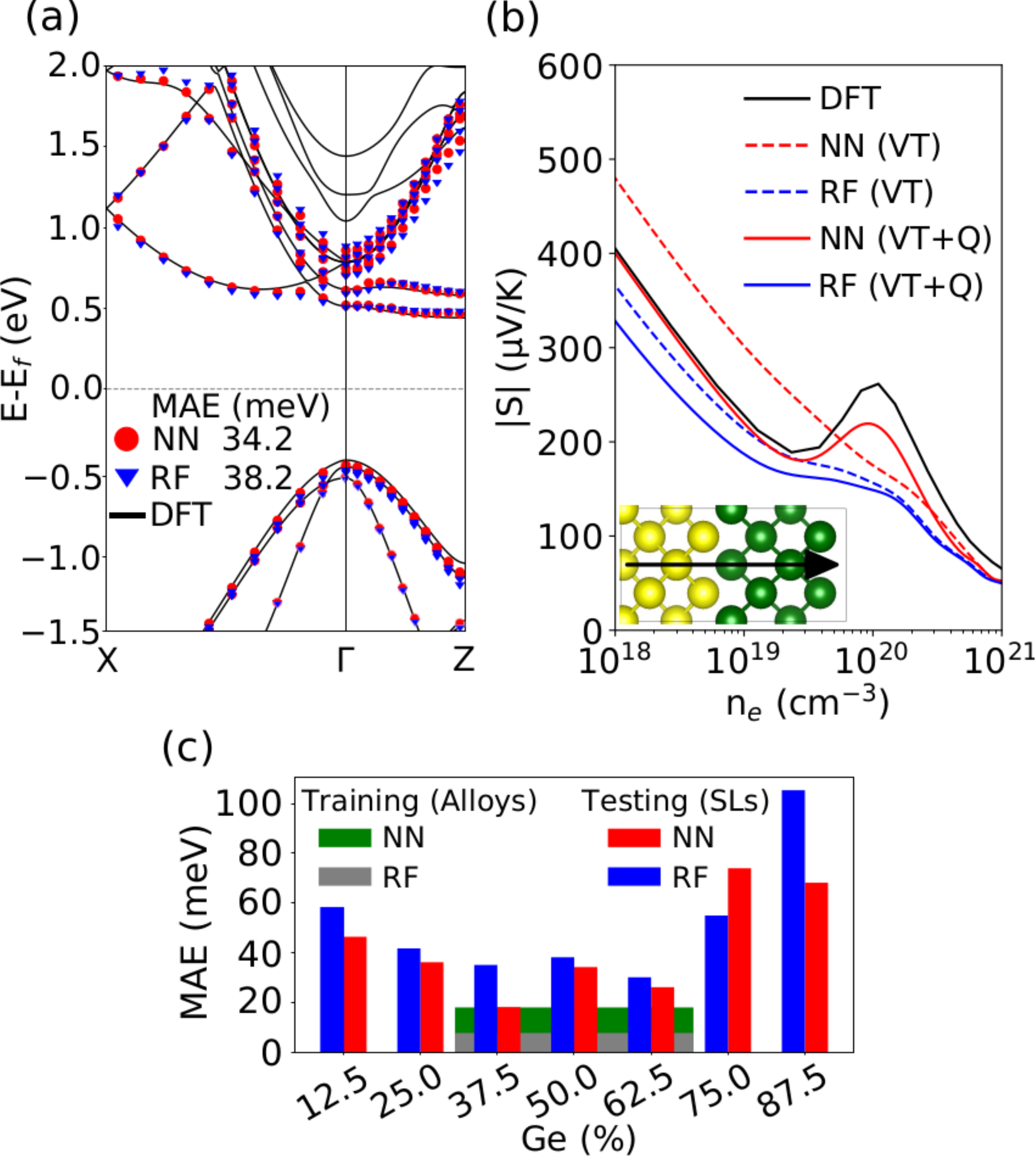}
\caption{{\bf Effectiveness of local structure-property preservation hypothesis:} (a) Energy bands of relaxed Si$_4$Ge$_4$ SL predicted with ML algorithms and compared with DFT results. {\bf ML model is trained on 350 disordered units, including local order parameter features ($Q$), and tested on 7 relaxed ordered SL structures.} (b) Thermopower obtained with DFT (black) and ML algorithms, NN (red) and RF (blue), trained with (solid) and without (dashed) order parameters, respectively. Improved match upon including order parameter features can be noted. (c) Train and test MAE of ML predicted energies.}
\label{fig:Q_importance}
\end{center}
\end{figure}

\subsection{Strained Ideal Superlattices}

We test the effectiveness of our ETI framework in predicting the thermopowers of ideal SLs, considered to be grown on substrates inducing epitaxial strain. We use the term ideal to refer to SLs with sharp interfaces. We consider 7 applied strain values ranging uniformly from $-1.1\%$ to $+6.1\%$, resulting in 49 different SLs, depicted by cartoons in Fig.~\ref{fig:strained_ideal_SL}(a). Strains in the range of $\sim3-4\%$ have been observed in Si/Ge nanowire heterostructures with compositionally abrupt interfaces, grown via the VLS process~\cite{wen2015strain}. We consider some extreme strains to probe the predictive power of our ML models. The models are trained on 40 and tested on 9 SLs. In Fig.~\ref{fig:strained_ideal_SL}(b), we show the bands of a strain-symmetrized Si$_4$Ge$_4$ SL along symmetry directions of a tetragonal BZ. Both NN and RF algorithms predict energies remarkably close to DFT results, with MAEs given by 13.2 meV and 27.0 meV, respectively. The train and test MAE for the two predictions are shown in Fig.~\ref{fig:strained_ideal_SL}(d,e). MAE is relatively small for small strain systems and higher for high strain boundary values. Both the algorithms yield small train MAE while their testing errors are considerably different. For example, the NN-predicted degenerate bands at $\sim$0.8 eV along $\Gamma-Z$ compare well with DFT results but the RF predictions deviate moderately. The band gap is also predicted better by the NN algorithm. In Fig.~\ref{fig:strained_ideal_SL}(c), we show $S$ of strain-symmetrized $n$-type Si$_4$Ge$_4$ SLs as a function of carrier concentration, $n_e$, which can be controlled by chemical or electrostatic doping methods~\cite{gupta2017electrostatic}. Within BTE, $S$ is obtained by integrating a function including band energies, Fermi-Dirac distribution function and transport distribution function~\cite{mahan1996best} over the respective BZ, as outlined in the Method section. Thus, the discrepancy in predicted bands leads to an accumulated error in $S$ prediction. The closer match of the NN-predicted lowest conduction bands with the DFT results in a better prediction of the resulting $S$. Figure~\ref{fig:strained_ideal_SL}(c) shows that the predictions significantly improve when the ML models are trained using VT-derived features (solid curves) in addition to using only global features (dashed curves). This analysis emphasizes the importance of considering local environment features in order to predict thermopowers with higher accuracy. 
\begin{figure*}
\begin{center}
\includegraphics[width=1.0\linewidth]{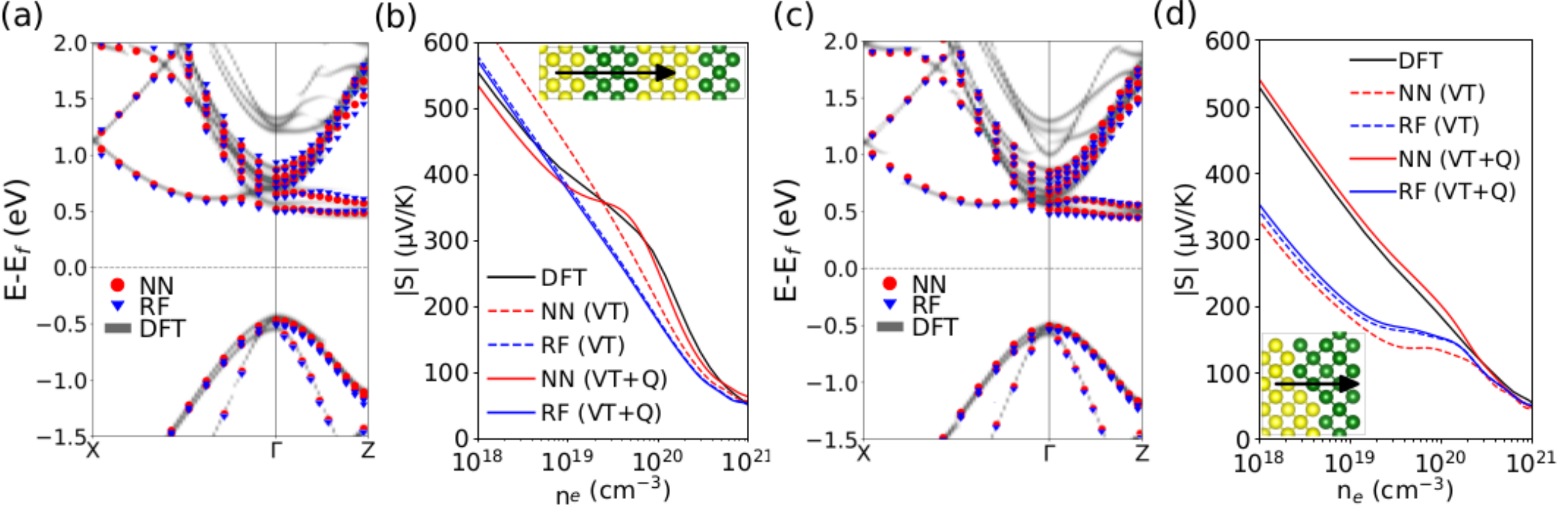}
\caption{{\bf Performance of ML algorithms on non-ideal test heterostructures:} (a) Effective band structure (EBS) of Si$_4$Ge$_4$Si$_5$Ge$_3$ multilayered system compared with ML predicted bands. (b) Thermopower calculated from DFT (black) and ML bands (NN-red and RF-blue). Solid and dashed lines represent ML predictions made with and without the knowledge of order parameters, respectively. (c) EBS of an imperfect layer heterostructure compared with ML predicted bands. (d) Thermopower calculated from DFT (black) and ML bands (NN-red and RF-blue). Insets show representative test system configurations.}
\label{fig:non_ideal_SL}
\end{center}
\end{figure*}

In the following, we provide further discussion on the effectiveness of training ML models with features describing local atomic environments. In Fig.~\ref{fig:Q_importance}(a,b), we show the bands of a strain-symmetrized Si$_4$Ge$_4$ SL along with the corresponding $S$. Similar to Fig.~\ref{fig:strained_ideal_SL}(b), the predictions match DFT results closely, with MAEs of 34.2 meV (NN) and 38.2 meV (RF), respectively. The remarkable aspect of these results is that the {\em ML models are trained only on disordered fragment units and the predictions are made for ordered structures.} These results provide a direct demonstration of our central hypothesis that the local atomic configurations-energy states relationship, $f(CN({\bf r},E)$, is preserved across configurations with different compositions. Figure~\ref{fig:Q_importance}(b) further highlights that training ML models including order parameter features improves $S$ predictions (solid curves). The MAE for the 7 relaxed SL configurations of varying Ge concentrations are shown in Fig.~\ref{fig:Q_importance}(c). The high MAE for the samples with the lowest and the highest Ge concentrations can be attributed to the limited number of disordered training units with similar Ge concentrations (see Fig~\ref{fig:crystal-graph}(e)). The order parameter maps thus provide great insight into the expected performance of the ML models for different test structures. These results demonstrate that our ML models capture the necessary information regarding the true interatomic interactions present in these binary heterostructures in an unbiased manner. We leverage this knowledge and the central hypothesis to probe $f(CN({\bf r}),E)$ in 16-atom ordered and disordered fragment units, and extrapolate the insight to predict the energy states and transport coefficients of larger heterostructures as demonstrated below. We train the ML models with both global and VT-derived features to achieve this objective.

\subsection{Non-Ideal Heterostructures}

To prove the transferability of the ETI framework, we task our ML models to predict electronic transport properties of 32-atom non-ideal SLs. The two types of ``non-idealities" we probe are represented by SLs with irregular layer thicknesses (Fig.~\ref{fig:non_ideal_SL}(b)), and imperfect layers (Fig.~\ref{fig:non_ideal_SL}(d)). These systems are explorable with first-principles techniques, but larger in size compared to the 16-atom training units. As a result, we face a challenge to validate the ML predicted bands against DFT results due to the size difference between train and test structures. The ML models predict energy bands sampling the first BZ of 16-atom models, as shown in Fig.~\ref{fig:strained_ideal_SL} and~\ref{fig:Q_importance}. However, the 32-atom test systems have a smaller BZ and as a result, several bands are zone-folded. Additionally, as the system size increases, so does the number of bands in both valence and conduction zones, making it challenging to keep track of. We resort to a band structure unfolding technique that allows to identify effective band structures (EBS), by projecting onto a chosen reference BZ~\cite{popescu2010effective, popescu2012extracting}. We obtain the EBS of different 32-atom test configurations by projecting the DFT computed bands onto the BZ of 16-atom reference cells, and compare with the ML predicted bands sampling a similar size BZ. This technique has been illustrated for different random substitutional alloy compositions: to probe to which extent band characteristics are preserved at different band indices, and $k$-points, compared to the respective bulk systems (see Methods section for details). Although this technique has not been applied to probe SL bands, we argue that these systems, especially non-ideal SLs, are closer to alloy systems due to broken translational symmetry. In Fig.~\ref{fig:non_ideal_SL}(a), we show the EBS of a 32-atom random multilayered heterostructure, Si$_4$Ge$_4$Si$_5$Ge$_3$. Here the indices represent the number of MLs in each component layers, as can be identified from the configuration in the inset of Fig.~\ref{fig:non_ideal_SL}(b). Figure~\ref{fig:non_ideal_SL}(c) shows the EBS of a 32-atom imperfect layer heterostructure, as represented by the configuration in the inset of Fig.~\ref{fig:non_ideal_SL}(d). The remarkable agreement between ML-predicted bands and EBS can be noted from both the figures. Similar to the example shown in Fig.~\ref{fig:strained_ideal_SL}, the NN algorithm provides a slightly better estimate of band gap. As demonstrated in Fig.~\ref{fig:non_ideal_SL}(b,d), the inclusion of the order parameters ($Q$) is crucial for accurate prediction of thermopower. We tested the ML models on a class of such non-ideal heterostructures and include other results in SI (see Fig.~S11).

\begin{figure}
\begin{center}
\includegraphics[width=0.95\linewidth]{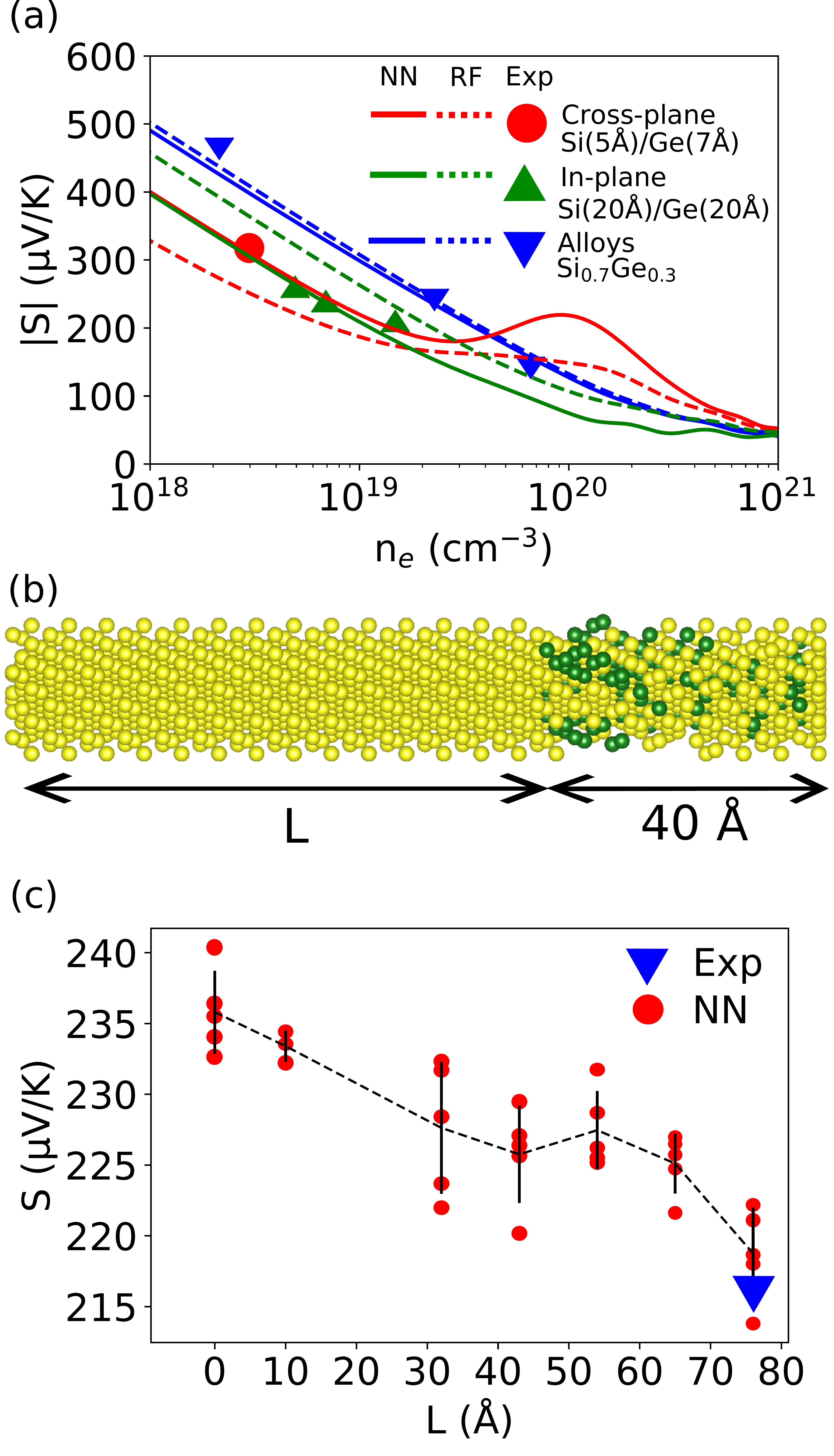}
\caption{{\bf Prediction of thermopowers of experimental systems by training ML models with first-principles modeling data obtained from 16-atom fragment units:} (a) Predicted thermopower (NN (solid lines), RF (dashed lines)) of $n$-type SLs and alloys compared to experimental data, taken from Ref.~\cite{koga2000experimental,yang2001characterization,dismukes1964thermal}. (b) Representative configuration of Si/Si$_{0.7}$Ge$_{0.3}$ SL. Length ($L$) of Si region is varied keeping length of Si$_{0.7}$Ge$_{0.3}$ region fixed. (c) $S$ of $p$-type Si/Si$_{0.7}$Ge$_{0.3}$ SLs at carrier concentration, $n_e = 1.5 \times 10^{19}$ cm$^{-3}$ predicted with NN algorithm and compared with experimental data~\cite{zhang2002measurement}. Predicted $S$ converge to measured value with increasing $L$ from below. Spread in NN predictions represent five randomized Si$_{0.7}$Ge$_{0.3}$ test configurations considered for each $L$.}
\label{fig:experimental_structures}
\end{center}
\end{figure}

\subsection{Experimental Heterostructures}

As we discussed previously, the domain of application of first-principles approaches is often limited to ideal systems that do not reflect the structural complexity of experimental heterostructure, mainly due to the required computational expenses. As a consequence, we resort to parametric approaches to predict electronic properties of experimental systems. It is highly desirable to establish a bridge between the domains of ({\em A}) {\em ab initio} accessible ideal systems and ({\em B}) experimental systems realized with nanofabrication techniques, to acquire parameter-free predictions of electronic properties of real systems. Our training units, as represented by cartoons in Fig.~\ref{fig:ETI-model}(a), fall within domain ({\em A}). A most challenging task faced by data-driven approaches arises in scenarios when the ML models are tested on cases that fall outside the domain of prior data. Below, we demonstrate that our ETI framework successfully predicts electronic properties of test systems representing domain ({\em B}), and thus establishes a bridge between the two domains. 

In Fig.~\ref{fig:experimental_structures}, we demonstrate the agreement between ML predicted thermopowers (solid (NN) and dashed (RF)) and measured values (circle and triangles) taken from Ref.~\citenum{yang2001characterization,dismukes1964thermal,koga2000experimental}. We chose three system categories to demonstrate the extrapolating power of our ETI framework: $n$-type Si/Ge SLs~\cite{koga2000experimental,yang2001characterization}, $n$-type SiGe alloys~\cite{dismukes1964thermal} and $p$-type Si/SiGe SLs~\cite{zhang2002measurement}. The triangles (green) in Fig.~\ref{fig:experimental_structures}(a) represent in-plane thermopowers of $n$-type Si(20\AA)/Ge(20\AA) SL grown along [001] direction at 300K~\cite{koga2000experimental}. We construct our model including 112 Si and 112 Ge atoms, relax the geometry as described in the Methods section, and extract features to obtain the ML prediction. The circle (red) in Fig.~\ref{fig:experimental_structures}(a) represents cross-plane thermopower of $n$-type Si(5\AA)/Ge(7\AA) SL grown along [001] direction at 300K~\cite{yang2001characterization}. We extract features from a model Si/Ge SL with 8 Si and 8 Ge atoms to acquire the prediction. The inverted triangles (blue) represent thermopowers of $n$-type Si$_{0.7}$Ge$_{0.3}$ alloys at 300K~\cite{dismukes1964thermal}, which we model by substituting 19 Si atoms with Ge in a 64-atom bulk Si supercell. The ML predictions show a good agreement for both cross-plane and in-plane thermopowers across all different carrier concentrations. The small deviations between ML results and experimental data can be attributed to the differences between local environments in the models and the experimental samples. We anticipate that the error in ML prediction would fall within experimental uncertainties. This comparison also reveals that ML predictions can be utilized to optimize the thermopowers of these systems by varying carrier concentrations. 

In Fig.~\ref{fig:experimental_structures}(c), we further establish that the ETI framework can guide heterostructure design with optimized electronic transport properties. We show the NN predicted cross-plane thermopowers of $p$-type Si/SiGe SLs at a carrier concentration $n_e = 1.5 \times 10^{19}$ cm$^{-3}$, as a function of varying Si layer thickness ($L$). A representative configuration of a Si/Si$_{0.7}$Ge$_{0.3}$ SL is shown in Fig.~\ref{fig:experimental_structures}(b). We model systems with a constant width alloy region and varied $L$. As can be noted from the figure that our predictions approach the experimental data obtained for the Si(80\AA)/Si$_{0.7}$Ge$_{0.3}$(40\AA) SL grown on a Si substrate~\cite{zhang2002measurement} as we approach $L\sim$ 80\AA. For each system with a given $L$, the spread in ML data refers to five models with different randomized substitutional alloy configurations. Our results establish the remarkable extrapolating power of the framework and also reveal that thermopower of Si/SiGe SLs can be optimized by choosing an appropriate system size guided by ML prediction. We argue that the extension of the prediction domain is enabled by our central hypothesis that local environment-electronic states relationships are preserved across configurations with different compositions. Training ML models with these relationships allows us to predict electronic transport properties of experimental heterostructures. This physics-based extrapolation is thus possible because of accumulating knowledge from ``known" environments.

\subsection{Scalability of ETI Framework}

\begin{figure}
\begin{center}
\includegraphics[width=1.0\linewidth]{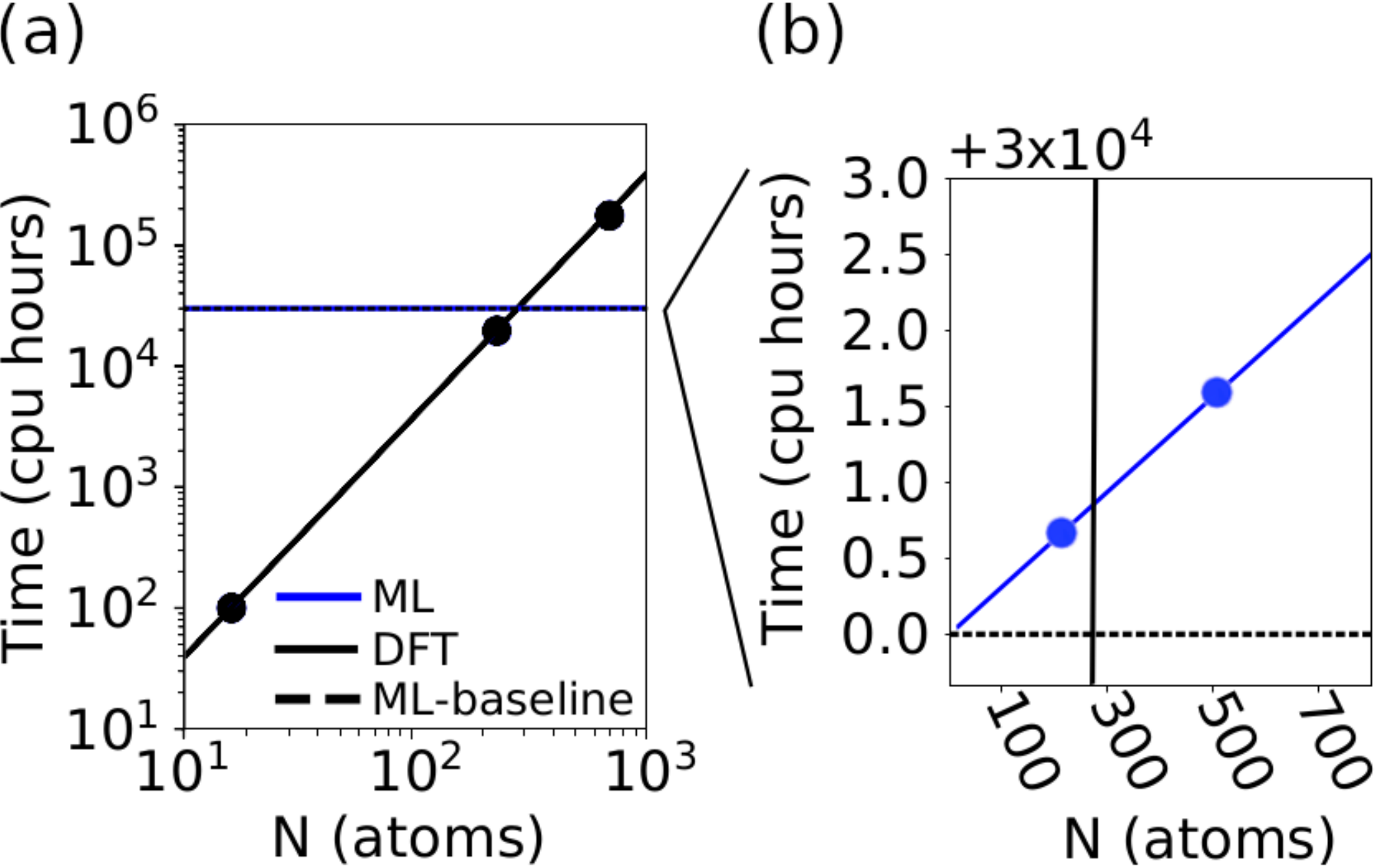}
\caption{{\bf Comparison of computational costs of DFT and ML approaches:} Time for DFT modeling scales quadratically with system size (solid black line). Time to obtain descriptors scales linearly with system size (solid blue line). Black dashed line represents baseline time to prepare training data.}
\label{fig:scaling_of_ETI}
\end{center}
\end{figure}

In order to further establish the claim that our ETI framework will help bridge the gap between {\em ab initio} accessible and fabricated systems, we explore the scalability of our framework with increasing system size. In Fig.~\ref{fig:scaling_of_ETI}, we compare the computational cost of using ETI model to predict electronic properties with direct DFT calculations, with increasing system size. The ML runtime is divided in two parts: the constant baseline, shown by the dashed line in inset of Fig.~\ref{fig:scaling_of_ETI}(b) refers to generation of training data with DFT; and the rest is devoted to feature extraction of DFT-relaxed test configurations. The plot shows that runtime for DFT calculations scales as $\sim N^2$ while that for feature extraction scales linearly with $N$, where $N$ is number of atoms. Figure~\ref{fig:scaling_of_ETI} establishes the remarkable advantage of the ETI framework for parameter-free prediction of thermopowers of large structures that can not be fully explored with DFT. We acknowledge that identifying the upper bound of this plot would be beneficial but leave it for future work.  

In summary, we demonstrate that the problem of predicting electronic properties of technologically relevant heterostructures can be largely solved by combining first-principles methods with ML techniques into a common framework. We illustrate that ML models are capable of extracting information regarding the true interatomic interactions present in ordered (layered) /disordered (alloy) semiconductor structures as a function of variable atomic environment, from the large body of atomistic data generated with individual DFT calculations. We train the ML models on the local environment-electronic state relationships in few-atom fragment units of varied atomic environments, and task the models to predict the environment-state relationships that develop in larger nanostructures, hosting similar local environments. The ML models show remarkable success in predicting thermopowers of experimental heterostructures, validated against experimental data. Our ETI framework establishes a bridge between ideal systems accessible with first-principles approaches and real systems realized with nanofabrication techniques. We elucidate that the extension of the prediction domain is facilitated by a central hypothesis that the local environment-electronic states relationship is preserved across configurations with different global compositions. Our prescription for electronic transport property prediction through codifying correlations in DFT generated electronic structure property data, breaks from previously developed methods that pursued a combinatorics-inspired optimization strategy, especially in the context of thermoelectric materials. Our viewpoint provides a path to extract important physics that determines transport properties of heterostructures, and allows to extend the applicability of first-principles techniques for technologically relevant heterostructures. We anticipate that this viewpoint would give our approach broad applicability to diverse materials classes.

\section{Methods}

\subsection{Training and Testing System Details}

We model Si$_n$Ge$_m$ SLs and ``alloys" with different compositions to generate the training data, where $n$ and $m$ refer to the number of Si and Ge atoms, respectively. We create a fragment unit Si$_n$Ge$_m$ ($n+m=16$) supercell by replicating a 8-atom conventional Si unit cell (CC) twice along the symmetry direction [001] and replacing $m$ Si atoms with Ge atoms, since both Si and Ge have stable FCC diamond lattice structures~\cite{book1, book2}. By replacing Si monolayers (ML) with Ge, we obtain 7 Si$_{8-x}$Ge$_x$ SLs, where $x$ is the number of MLs: $x=[1,2, \dots 7]$. The disordered SiGe structures are prepared with similar 16 atom supercells, two conventional 8-atom cells (CC) stacked along the [001] direction. For each chosen Ge concentration $(5/16,\; 6/16,\; 7/16,\; 8/16,\; 9/16,\; 10/16,\; 11/16)$, we generate 50 substitutional ``alloy" configurations, resulting in 350 total disordered fragment training units. To generate train and test strained SL structures, we consider applied strains ranging uniformly from $-1.1\%$ to $+6.1\%$ with a total of 7 different strain values, resulting in 49 different SLs, as shown in Fig.~\ref{fig:strained_ideal_SL}. We estimate the in-plane strain in the SLs from the lattice constants by $\epsilon_{\parallel}= (a_{\parallel}/a_{\text{Si}}-1)$~\cite{van1986theoretical} with $a_{\text{Si}}=5.475\;\text{\AA}$. 

The non-ideal heterostructures shown in Fig.~\ref{fig:non_ideal_SL}(b) and Fig.~\ref{fig:non_ideal_SL}(d) are modeled with 32 atom supercells (4 CCs). We construct the model Si(20\AA)/Ge(20\AA) SL from $2 \times 2 \times 7$ CCs including 112 Si and 112 Ge atoms and relax the geometry without any applied constraints. We model the Si(5\AA)/Ge(7\AA) SL shown in Fig.~\ref{fig:experimental_structures}(a) with $1 \times 1 \times 2$ CCs that include 8 Si and 8 Ge atoms. The Si$_{0.7}$Ge$_{0.3}$ alloy is modelled using a randomly substituted 64-atom $2 \times 2 \times 2$  CCs system that includes 45 Si and 19 Ge atoms. The experimental structures shown in Fig.~\ref{fig:experimental_structures}(b) are modeled with a SiGe random alloy region ($2\times2\times7$ CCs) and a Si layer of varied lengths between 0 and 10 CCs. We model systems with total size varied from $2 \times 2 \times 7$ (224 atoms) to $2 \times 2 \times 17$ CCs (544 atoms) by increasing $L$ and keeping the width of the alloy region constant to $2 \times 2 \times 7$ CCs. For each system with a given $L$, we model the substitutional alloy region with five different randomized configurations.

\subsection{DFT Computational Details}

The relaxed geometries of the structures are obtained using Vienna Ab Initio Simulation Package (VASP) package. The lattice constants and the atomic positions in Si$_n$Ge$_m$ structures are optimized using Broyden-Fletcher-Goldfarb-Shanno Quasi-Newton algorithm, sampling the Brillouin zone (BZ) with $8\times8\times 8$ $k$-point mesh. To simulate SLs under applied strain, we keep the cell volume fixed and relax the cell shape in every direction.
We perform the electronic structure calculations with DFT using the generalized gradient approximation (GGA) implemented in the Vienna Ab initio Simulation Package (VASP) \cite{kresse1996efficiency,kresse1996efficient} with the Perdew-Burke-Ernzenhof (PBE) exchange-correlation functional~\cite{perdew1996generalized}. The ultra-soft projector-augmented wave (PAW) pseudopotential~\cite{kresse1999ultrasoft,blochl1994projector} with a cutoff energy of 400 eV was used to describe the interaction between the valence electrons and the ions. For the self-consistent calculations, the energy convergence threshold was set to $10^{-6}$ eV. We haven't included spin-orbit interaction in our analysis since the magnitude of the lattice strain induced splittings is larger than the spin-orbit splittings~\cite{hybertsen1987theory}. The electronic bands are plotted along the $\Gamma-Z$ symmetry direction of the BZ with 11 points resolution. Following relaxation, we perform non self-consistent field (NSCF) calculations to obtain the band energies using a dense $\Gamma$-centered $21\times 21\times 21$ Monkhorst-Pack $k$-point mesh~\cite{monkhorst1976special}, to sample the irreducible Brillouin zone (IBZ). Such sampling is necessary to converge the calculation of the electronic transport coefficients. Once the electronic structure calculations are completed, we employ the semi-classical Boltzmann transport theory~\cite{ziman1960electrons} as implemented in BoltzTraP code~\cite{madsen2006boltztrap} to compute the room temperature Seebeck coefficients. The $k$-point mesh is chosen after systematic studies to converge Seebeck coefficients with increasing the mesh size. In Fig.~S8 and~S9, we show the convergence of $S$ of two representative configurations with increasing the $k$-sampling and number of included bands, respectively. 

\subsection{Effective Band Structures}

\noindent Following the approach outlined in Ref.~\citenum{popescu2012extracting}, we transform the band structures of larger configurations into EBS of a reference cell consisting of 16 atoms, using spectral decomposition~\cite{wang1998majority}. The reference cell contains the same number of atoms as the training units and is approximately of the same size as 2 CCs stacked along [001] direction. However, the dimensions of the reference cells that each test configuration is projected to are different, and are obtained by dividing the supercells as multiples of 2 CCs and taking an average. We calculate the eigenstates $|\overrightarrow{K} m\rangle$ of the test supercells using DFT, sampling the BZ with a $21\times21\times21$ $K$-point mesh, where $m$ is the band index. The spectral weight that quantifies the amount of character of Bloch states $|\overrightarrow{k_i} n\rangle$ of the reference unit cell preserved in $|\overrightarrow{K} m\rangle$ at the
same energy $E_m = E_n$, can be written as 
\begin{equation}
P_{\overrightarrow{K} m}(\overrightarrow{k_i})=\sum_n | \langle \overrightarrow{K} m |\overrightarrow{k_i} n \rangle|^2.
\label{eq:SpectralWeight}
\end{equation} 
The spectral function (SF) can then be defined as
\begin{equation}
A(\overrightarrow{k_i},E)=\sum_m P_{\overrightarrow{K} m}(\overrightarrow{k_i}) \delta (E_m-E),
\label{eq:SpectralFunction}
\end{equation} 
where $E$ is a continuous variable of a chosen range over which we probe for the preservation of the Bloch character of the supercell eigenstates. The delta function in Eq.~\ref{eq:SpectralFunction} is modeled with a Lorentzian function with width 0.002 eV. $A(\overrightarrow{k_i},E)$ are normalized by dividing the spectral functions by $max_{\{\overrightarrow{k_i},E\}}[A(\overrightarrow{k_i},E)]$.

\subsection{Seebeck Coefficients}

We compute the Seebeck coefficients using the semi-classical BTE as implemented in the BoltzTraP code~\cite{madsen2006boltztrap}. All thermopower calculations are performed at room temperature and for technologically relevant high doping regime ranging from n$_e = 10^{18}$ to $10^{21}$ cm$^{-3}$. $S$ is obtained from $(1/eT)(\mathcal{L}^{(1)}/\mathcal{L}^{(0)})$, where $e$ is the electron charge, $T$ is temperature, and the generalized in-plane ($\parallel$) or cross-plane ($\perp$) $n^{\text{th}}$-order conductivity moments are, 
\begin{equation}
\mathcal{L}_{\parallel, \perp}^{(n)}(\epsilon_F, T)=\int d\epsilon \Sigma_{\parallel, \perp} (\epsilon) (\epsilon-\epsilon_F)^{(n)}\left(-\frac{\partial f}{\partial \epsilon} \right). 
\label{eq:Lintegral}
\end{equation} 
The integrand is computed from the energy difference ($\epsilon-\epsilon_F$) to the $n^{\text{th}}$ power, the Fermi energy level ($\epsilon_F$), the derivative of the Fermi-Dirac distribution function ($f$) with respect to energy $\epsilon$, and the transport distribution function (TDF)~\cite{mahan1996best}. TDF can be expressed as
\begin{equation}
\Sigma_{\parallel, \perp} (\epsilon) = \frac{\tau}{\hbar (2\pi)^3} \oint_{\epsilon_{\bf k}=\epsilon} \frac{d\mathcal{A}}{|{\bf v_k}|} ({\bf v}_{{\bf k},(\parallel, \perp)})^2,
\label{eq:sigma}
\end{equation}
within the constant relaxation time ($\tau$) approximation. The area-integral is given by the DOS ($\propto \oint_{\epsilon_{\bf k}=\epsilon} \frac{d\mathcal{A}}{|{\bf v_k}|}$) weighted by the squared group velocities, $({\bf v}_{{\bf k},(\parallel, \perp)})^2$.

It is known that the PBE-GGA approach poorly predicts semiconductor band gaps~\cite{perdew2017understanding,morales2017empirical}, as opposed to using hybrid functionals~\cite{hummer2009heyd}. Nevertheless, the PBE-GGA approximation has been regularly employed to compute the electron/hole transport coefficients of semiconductors, including thermoelectric properties of [111]-oriented Si/Ge SLs~\cite{hinsche2012thermoelectric}. These studies demonstrate the effectiveness of the PBE-GGA approximation to depict the role of lattice environment on electronic properties of Si-based systems. In previous publications, we discussed the discrepancy in bandgap predictions in detail~\cite{proshchenko2019optimization} as well as shown comparisons of $S$ of Si$_4$Ge$_4$ SLs predicted using the Heyd-Scuseria-Ernzerhof~\cite{heyd2003hybrid} and the PBE functionals~\cite{proshchenko2019modulation}. We find that the PBE-predicted $S$ vs n$_e$ relationship closely follows the HSE prediction for low strain cases, and shows small deviations at low doping concentrations for high strain cases, which can be attributed to bandgap discrepancies~\cite{proshchenko2019modulation}. In addition, we tested that using a scissors operator for band gap correction using the HSE predicted gaps (See Ref.~\citenum{proshchenko2019modulation}) or experimental band gap (Fig.~S8), essentially leaves the $S$ vs n$_e$ curve unchanged. This systematic analysis showed the robustness of our results highlighting the relationship between lattice environment and electronic transport in heterostructures, independent of the numerical approach used, and motivated us to use PBE-GGA-BTE approach to analyze the thermopowers of Si$_n$Ge$_m$ heterostructures. In the present article, we use a static correction ($U_{\text{GGA}}=0.52$ eV~\cite{hinsche2012thermoelectric}) to match the PBE predicted band gap to the measured band gap value for bulk silicon. The PBE approach is especially suited for data driven studies since it is far less expensive compared to a more accurate hybrid functional. For example, the electronic bands calculation of a Si$_4$Ge$_4$ SL using PBE, over a $21\times21\times21$ $k$-point mesh, required 31 CPU hours and compared to 1075 hours of CPU time when using the hybrid functional. 

We used a constant relaxation time ($\tau$) approximation for all the calculations presented in this manuscript. This approximation allows us to calculate $S$ without any free parameters. It is a common practice to obtain $\tau$ by fitting experimental mobility data for specific carrier concentrations with empirical approximations, and adjust the first-principle results accordingly to reproduce experimental findings. For example, the first-principles estimation of electronic transport properties of strained bulk Si used relaxation times fitted from the measured mobility data of unstrained Si~\cite{hinsche2012thermoelectric}. One main reason is that first-principles computation of $\tau$ is highly expensive for model systems containing greater than a few atoms. As a result, only a handful of previous studies exist that analyzed the electronic properties of highly technologically relevant Si/Ge heterostructures using first-principle methods, especially with including the complex effects of strain or non-idealities. It is known that strain could alter the dominant scattering processes in bulk Si~\cite{dziekan2007theoretical}, however, the role of different scattering mechanisms on electron relaxation in Si/Ge heterostructures due to lattice strain or defects is relatively unexplored. In an earlier publication, we estimated the relaxation time assuming that the electron-phonon scattering rates in non-polar semiconductors generally are proportional to the density of states (DOS), and provided a comparison between $S$ computed with constant $\tau$ and with $\tau(\epsilon)\propto 1/\text{DOS}$~\cite{proshchenko2019optimization}. We noted that $S$ trends match quite well between the two approximations, although the exact values differ. These observations motivate us to follow a similar approach to compute the electronic transport coefficients in this article. We acknowledge that a detailed analysis of the validity of this approximation would be highly beneficial. However, such a study is out of scope of the present manuscript, especially since our test systems include 100s of atoms. Our aim here is to establish that the local functional relationships present in small models can be harnessed to achieve parameter free prediction of the electronic transport properties of experimental heterostructures. And we have provided a proof of concept by demonstrating that our predictions, made using a constant relaxation time, match the measured data.

\section{Data availability}
The authors declare that the data supporting the findings of this study are available within the main article and the Supplementary Information document.

\section{Acknowledgements}
The project is funded by the Defense Advanced Research Projects Agency (Defense Sciences Office) [Agreement No.: HR0011-16-2-0043]. This work utilized resources from the University of Colorado Boulder Research Computing Group, which is supported by the National Science Foundation (awards ACI-1532235 and ACI-1532236), the University of Colorado Boulder, and Colorado State University. This work used the Extreme Science and Engineering Discovery Environment (XSEDE), which is supported by National Science Foundation grant number ACI-1548562.  

\section{Author contributions}

\noindent A.K.P contributed to the acquisition and the analysis of data and the creation of new scripts used in the study; S.N. contributed to the conception and the design of the work, the interpretation of data, drafting and revision of the manuscript.

\section{Competing interests}

\noindent The authors declare no competing interests.

\bibliography{MLLiterature}

\end{document}